\documentclass[aps,prb,twocolumn,groupedaddress]{revtex4}
\usepackage{times,xspace}
\usepackage{amsbsy,amssymb,amsmath,bm}
\usepackage{color,epsfig,rotate}
\usepackage{amsfonts}
\usepackage{amsmath}
\usepackage{amssymb}
\begin{document}
\title{Electronic Orbital Currents and Polarization in Mott Insulators}
\author{L.N. Bulaevskii and C.D. Batista}
\affiliation{Los Alamos National Laboratory, Los Alamos, New Mexico 87545}
\author{M. Mostovoy}
\affiliation{ Groningen University, Nijenborgh 4, 9747AG Groningen, The Netherlands }
\author{ D. Khomskii$^1$}
\affiliation{ Physikalisches Institut, Universit\"at zu K\"oln, Z\"ulpicher Strasse 77, D-50937 K\"oln, Germany}
\date{\today}

\begin{abstract}
The standard view is that at low energies Mott insulators exhibit only magnetic properties 
while charge degrees of freedom are frozen out as the electrons become localized 
by a strong Coulomb repulsion.
We demonstrate that this is in general not true: for certain spin textures 
{\it spontaneous circular electric currents} or {\it nonuniform charge distribution} exist in 
the ground state of Mott insulators. In addition, low-energy ``magnetic'' states contribute
comparably to the dielectric and magnetic functions $\epsilon_{ik}(\omega)$ and $\mu_{ik}(\omega)$
leading to interesting phenomena such as rotation 
the electric field polarization and resonances which may be common for both functions
producing a negative refraction index in a window of frequencies. 

\end{abstract}

\maketitle
Mott insulators are the paradigm of strongly correlated materials. 
Their minimal Hamiltonian is the Hubbard model which includes a hopping term, 
$t$, and on--site Coulomb interaction $U$.
At half filling (one electron per site) and in the large $U/t$ limit, each 
site is occupied by a single electron to avoid the strong on-site repulsion. 
The charge becomes localized by this mechanism and the low energy properties are described 
by the remaining spin  degrees of freedom. For this reason,
Mott insulators with large $U/t$ have been traditionally considered as materials which have 
only magnetic properties at low energies due to their spin moments.  
Despite this common conviction, we will show here that certain ground states 
of Mott insulators exhibit real electric currents in loops (orbital currents) that produce
orbital magnetic moments, while others show modulation of electron
charge (polarization). Consequently, spins in Mott insulators are coupled 
not only to dc magnetic fields, but also to dc electric fields, and it is possible to
have magnetically driven electronic ferroelectricity \cite{Tokura,Cheong,Khomskii,Batista}. 
Moreover, nonvanishing matrix elements of the polarization between the 
ground state and excited magnetic states result in a nonvanishing 
contribution  to $\epsilon_{ik}(\omega)$ at low energies 
with optical strengths comparable to those of $\mu_{ik}(\omega)$ for the diagonal and  off-diagonal 
elements. Therefore, rotation of the electric field polarization is a 
characteristic signature of spin textures with orbital currents.

The apparent contradiction between the insulating nature of the system and 
the existence of non-zero orbital currents is resolved when we notice that
electrons are not completely localized on their ions for finite $U/t$. 
In fact, the effective Heisenberg interaction, 
$J \propto t^2/U$, results from a partial delocalization: an electron
gains kinetic energy by ``visiting'' virtually a neighbouring site, but this only occurs
if the spins are opposite on both sites (Pauli principle). Simlarly, 
the electron moves along a closed loop generating local currents that 
depend on the spin structure along the loop. 
 
We start by considering a half filled Hubbard model on a general lattice:
\begin{equation}
H= -\sum_{ij \sigma} t_{ij} (c^{\dagger}_{i \sigma} c^{\;}_{j\sigma}
+c^{\dagger}_{j \sigma} c^{\;}_{i \sigma}) +
\frac{U}{2}\sum_i( n_{i}-1)^2,
\label{Hubbard}
\end{equation}
where sites are labeled by indices $i,j$, $c^{\dagger}_{i\sigma}$ ($c_{i\sigma}$) is the creation (annihilation) 
operator of an electron  with spin $\sigma$ on site $i$ and $n_i =\sum_{\sigma} c^{\dagger}_{i \sigma} c^{\;}_{i\sigma}$ 
is the number operator. 
The low energy spectrum of this model is described by an
effective Heisenberg spin Hamiltonian, ${\tilde H}$, which is obtained by the usual 
degenerate perturbation theory in $t/U<<1$. 
${\tilde H}$ acts on the low--energy subspace where all the sites are singly occupied. Consequently, 
it is expressed in terms of the spin operators: 
$S_i^{\eta}=\sum_{\mu,\nu} c^{\dagger}_{i \mu} \sigma^{\eta}_{\mu\nu}c^{\;}_{i \nu}$, where
$\sigma^{\eta}$ are the Pauli matrices and $\eta=\{x,y,z\}$.
The expression of $\tilde{H}$ to order $t^2$ is: 
$\tilde{H}^{(2)}= \sum_{ij} J_{ij}( {\bf S}_{i}\cdot {\bf S}_{j}-1/4)$,
with $J_{ij}=4t_{ij}^2/U$. In general, any physical operator, $O$, has an expression in terms of spin operators, $\tilde{O}$,
that results from the application of degenerate parturbation theory. 

We will consider first the current operator 
\begin{equation}
{\bf I}_{ij} = \frac{iet_{ij} {\bf r}_{ij}}{\hbar r_{ij}} \sum_{\sigma}( c^{\dagger}_{j\sigma} c^{\;}_{i\sigma} -
c^{\dagger}_{i\sigma} c^{\;}_{j\sigma})
\label{eq:cu}
\end{equation}
between sites $i$ and $j$. Since the shortest loop is a triangle, the lowest order 
finite contribution to the current operator is $t^3/U^2$ and contains the 
product of 3 spin operators. The current is a scalar under spin rotations and 
it is odd under time reversal and under spatial inversion. 
The only possible expression involving three spin operators is the 
so--called  ``scalar spin chirality'' operator.
Using perturbation theory \cite{Bogolyubov49}, we find that the contribution to the current 
in the bond $1,2$ from the triangle 1-2-3 is:
\begin{equation}
{\bf I}_{12,3}=\frac{{\bf r}_{12}}{r_{12}}~\frac{24e}{\hbar }\frac{t_{12}t_{23}t_{31}}{U^{2}}
\left[ \mathbf{S}_{1}\times \mathbf{S}_{2}\right] \cdot \mathbf{S}_{3}.
\label{eq:current}
\end{equation}
The quantity $\chi _{ij,k}=[\mathbf{S}_{i}\times 
\mathbf{S}_{j}]\cdot  \mathbf{S}_{k}$ called scalar spin chirality was introduced previously in 
numerous discussions of  magnetic systems. \cite{chirality} It was invoked in the theory of 
anyon superconductivity \cite{Wen} and for describing  properties of triangular and kagome magnets \cite{Momoi,Schweika}. 
Scalar chirality produces a novel Berry-phase mechanism for anomalous Hall effect \cite{Taguchi}.
Chirality can lead to new universality classes in phase transitions \cite{Kawamura98}
and  special ``chiral glass'' phases in disordered systems \cite{Kawamura2}. 
It can also modify quantum tunneling in magnetic molecules \cite{Nojiri} and 
it appears in the electronic contribution to the Raman scattering \cite{Shastry}.
Despite this broad interest, the physical meaning of scalar chirality remained unclear.
Eq.(\ref{eq:current}) shows that scalar spin chirality corresponds to orbital currents running in 
the low energy states of Mott insulators. These currents produce orbital magnetic moments  
$\tilde{{\bf L}}_{ijk} \propto \chi_{ij,k} {\hat {\bf z}}$, where ${\hat {\bf z}}$ is normal to 
the plane of the triangle. We note that orbital currents only appear for
noncoplanar spin structures, as $\tilde{L}_z$ is proportional to a solid angle
formed by the spin vectors $\mathbf{S}_{1}$, $\mathbf{S}_{2}$ and $\mathbf{S}_{3}$. 

\begin{figure}[!htb]
\vspace*{-0.8cm}
\hspace*{0.5cm}
\includegraphics[angle=90,width=0.4\textwidth]{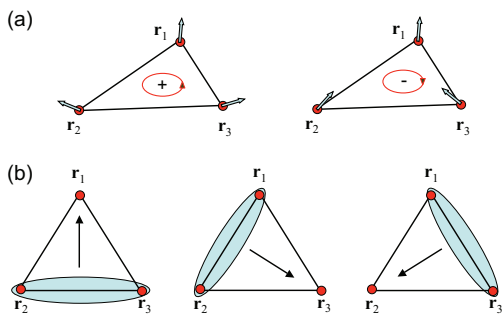}
\vspace{-0.6cm}
\caption{(a) Ground states with nonzero electric current of the $C_3$ invariant Heisenberg triangle.
The circular arrows indicate circular currents, and $(+, -)$ the sign of a scalar chirality.
(b) Examples of magnetic states with nonzero polarization. The two spins inside the oval form a singlet 
state. The unpaired spin can be up or down}
\label{triangle}
\end{figure}

Normally, orbital magnetism leads to a paramagnetic response. 
However, since the coupling of magnetic field to orbital moments is weak ($
\propto t^{3}/U^2$) compared to the spin Zeeman coupling, 
the dependence of $L_{z}$ on the external magnetic field is mainly due to 
changes in the spin configuration. For this reason $\tilde{L}_{z}$ can increase or decrease with field. 
For instance, the coplanar 120$^{\circ }$ spin structure  
in the easy-plane triangular lattice is turned into an ``umbrella'' pattern
when a magnetic field perpendicular to the easy-plane is applied, see Fig.~\ref{triangle}a. 
The orbital moment on a triangle $ijk$ becomes nonzero 
and its absolute value first increases with field but finally decreases to
zero when the spins become fully polarized. 

The orientation of the orbital moments depends 
on the signs of $t_{ij}$. For lattices, the current on a given bond is the sum of the loop currents
in all the triangles to which that bond belongs:
$\tilde{\bf I}_{ij} = ({\bf r}_{ij}/r_{ij}) \sum_{k} I_{ij,k}.$
In some particular cases, the net current of the bond is zero because
different contributions cancel each other.
In this situation a net current exists on the surface of
two--dimensional lattices. 

\begin{figure}[!htb]
\vspace*{-0.7cm}
\hspace*{0.0cm}
\includegraphics[angle=90,width=0.5\textwidth]{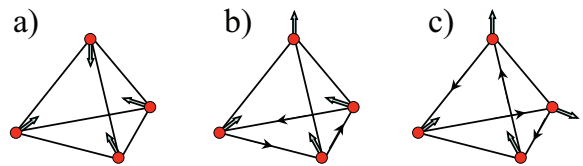}
\vspace{-4.5cm}
\caption{ Non--coplanar spin configuration in a pryrochlore lattice. a) Four 
spins point inwards along the principal diagonals; the net currents are zero. b) Three spins
point inwards while the other one points outwards leading to a net current circulating in the
opposite triangle. c) Two spins point inwards and the other two point outwards. The  
orbital current circulates in a loop formed by four edges of the tetrahedron.}
\label{tetra}
\end{figure}

\begin{figure}[!htb]
\vspace*{-0.3cm}
\hspace*{0.2cm}
\includegraphics[angle=0,width=0.4\textwidth]{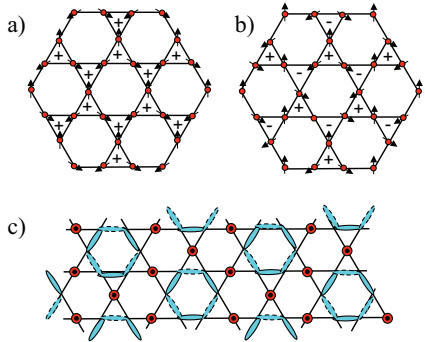}
\vspace{-0.7cm}
\caption{Spin configurations on a kagome lattice that lead to a current or charge ordering in an
applied magnetic field. a) The ``umbrella'' phase induced by a field perpendicular to the plane has
a uniform current ordering. The $\pm$ signs denote directions of current. b) The same as a) for staggered current ordering.
c) Spin ordering induced by field for a Heisenberg model on a kagome lattice \cite{Zhitomirsky05,Cabra05}. 
The elongated ovals indicate a resonant valence bond state on the corresponding hexagons, while the little circles 
represent spins that are polarized along the field direction. This structure is accompained
by charge ordering: the charges on sites belonging to hexagons and those on isolated sites should be different.}
\label{kagome}
\end{figure}

For instance, for regular spin tetrahedra (building blocks of pyrochlore or 
spinel lattices) with strong uniaxial [$\pm1,\pm1,\pm1$] anisotropy, the 
structures ``four in'' or ``four out'' do not have net currents, see Fig.\ref{tetra}a. 
However, the structures ``two in -- two out'' (spin ice), or ``three in -- one out'', which could be 
stabilized by an external field, have nonzero net 
orbital currents as shown in Figs.\ref{tetra}b,c. 
Similarly, two widely discussed structures in kagome lattices are those with a homogeneous vector chirality
${\bf S}_1\times{\bf S}_2+{\bf S}_2\times{\bf S}_3+{\bf S}_3\times{\bf S}_1$ (${\bf q}=0$ structure), 
and with staggered vector chirality ($\sqrt{3} \times \sqrt{3}$ 
structure) \cite{Schweika,Zhitomirsky05,Cabra05}. In both cases, 
there is easy plane anisotropy and the umbrella structure induced by a magnetic field perpendicular
to the lattice has a nonzero orbital moment. As shown in Fig.\ref{kagome}, the pattern of currents 
and orbital moments is uniform in the first case, and staggered for the latter case 
(despite the fact that the net spin moment is the same). The coupling of a net orbital moment 
to an external magnetic field favors the uniform state.


In a similar way we derive an expression for the projected local electron  number operator $\tilde{n}_i$. 
This operator is a scalar under rotations in spin--space, i.e., it must be a function of the combinations 
${\bf S}_{i} \cdot {\bf S}_{j}$. The first non--zero contribution to a 
deviation $\delta \tilde{n}_{i}$ from unity is 
\begin{equation}
\delta \tilde{n}_{1}= {\tilde n}_1-1 = 
8 \frac{t_{12}t_{23}t_{31}}{U^{3}}\left[ \mathbf{S}_{1}\cdot \left( 
\mathbf{S}_{2}+\mathbf{S}_{3}\right) -2\mathbf{S}_{2}\cdot \mathbf{S}_{3}
\right].
\label{eq:charge}
\end{equation} 

A similar expression holds for the charges at sites 2 and 3 after a cyclic permutation of indices.
The spin structure of the charge operator is uniquely fixed by the
invariance of $\tilde{n}_{1}$ under the time reversal symmetry and the interchange of 
sites 2 and 3, as well as by the conservation of total charge of the
triangle: $\sum_{i=1}^3 \delta \tilde{n}_i =0$. 
The redistribution of charges within triangles induces electric dipoles in
magnetic states: $\tilde{{\bf P}}_e = e\sum_{i} {\bf r}_i \delta\tilde{n}_i$. These dipoles can appear spontaneously 
or can be induced by a magnetic field, see Fig.\ref{triangle}b. 
In particular, a triangle with the classical coplanar 120$^{\circ }$
ordering of spins does not have charge redistribution in the ground state. However, for an easy-plane 
anisotropy, an in-plane magnetic field perpendicular to the bond 2-3 cants the spins in such a way that
$\langle \mathbf{S}_{1}\cdot \mathbf{S}_{2} \rangle=\langle \mathbf{S}_{1}\cdot \mathbf{S}_{3}\rangle$ becomes
larger than $\langle \mathbf{S}_{2}\cdot \mathbf{S}_{3}\rangle$ inducing an electric
dipole moment in the field direction. Thus, a nonzero electric polarization of pure electronic origin 
appears in such nonuniform spin configuration.

The projected dipole and current operators are
identically zero for {\it bipartite} lattices with nearest--neighbor hoppings \cite
{Lev67}. This results from the invariance of $H$ under the product of particle-hole and $t\rightarrow -t$
transformations, while ${\bf P}$ and ${\bf I}$ are odd under this transformation. 

We will discuss now some consequences of the obtained results.
The electric dipole induced by virtual electron hopping has the same form 
as the one resulting from the dependence of the exchange constants on 
ion displacements $\mathbf{u}_{i}$ in a magnetically ordered state (magnetostriction): 
$J_{ij}\approx J_{ij}(0) + \mathbf{u}_{n}\cdot \nabla _{n}J_{ij}$. 
Minimizing the sum of the magnetic energy $\sum J_{ij}\mathbf{S}_{i}\cdot \mathbf{S}_{j}$ and the lattice distortion
energy respect to $\mathbf{u}_{i}$, we find that the resulting electric dipole 
of a triangle is expressed in terms of scalar products 
$\mathbf{S}_{i}\cdot \mathbf{S}_{j}$. Due to the symmetry considerations discussed above, the spin structure of the dipole is
the same as in Eq.(\ref{eq:charge}), while the coefficient is $\sim
e\left\vert \nabla J\right\vert /K$, where $K$ is the lattice spring constant.
Electronic dipoles, $\tilde{{\bf P}}_e$, corresponding to Eq.~(\ref{eq:charge}), 
together with the spin-dependent dipoles originating from magnetostriction lead to
the coupling between spins and electric field ($eZ_l$ is the charge of an ion $l$):
\begin{equation}
\tilde{H}_e=-\tilde{{\bf P}}\cdot{\bf E}, \ \ \ {\tilde{\bf P}}=
{\tilde{\bf P}}_e+\sum_{ijl}eZ_lK_l^{-1}({\bf S}_i\cdot{\bf S}_j-1/4).
\end{equation}

A simple example of a spin-driven charge modulation is given by the  
the 1/3 plateau phase of the $S=1/2$ kagome lattice. This state has the local structure shown in Fig.\ref{kagome}c, 
with a resonating singlet state on the hexagons and up-spins in between \cite{Zhitomirsky05}, 
and could be a long-range ordered valence-bond 
crystal \cite{Cabra05}. A similar situation arises forthe several magnetization plateaus in the 
Shastry-Sutherland system SrCu$_2$(BO$_3$)$_2$ \cite{Shastry-Sutherland}. The states at each 
plateau consist of ordered arrays of singlet and triplet dimers which according to (\ref{eq:charge}) 
should lead to a spin driven charge density wave. There are also systems consisting of isolated triangles 
with long-range magnetic ordering, e.g. La$_4$Cu$_3$MoO$_{12}$ \cite{Broholm}. 
According to Eq.~(\ref{eq:charge}), the magnetic structure found in  Ref.~\onlinecite{Broholm} should  
have nonzero total electric polarization (multiferroic behavior).

Next we consider the response of the Mott insulator to ac electric and magnetic fields. 
The matrix elements of $\tilde{{\bf P}}$ between the ground state $|0\rangle$ and excited magnetic 
states $|n\rangle$ define the contribution of these states to the dielectric function,
\begin{equation}
\epsilon_{ik}(\omega)=\epsilon_0\delta_{ik}+\frac{8\pi}{V}\sum_n\frac{\omega_{no}
\langle 0|{\tilde P}_i|n\rangle \langle n|{\tilde P}_k|0\rangle} 
{(\omega^2-\omega_{n0}^2-i\delta)}, \label{ep}
\end{equation}
at $T=0$ and at frequencies well below the frequencies of optical phonons. Here $\delta\rightarrow 0$,
$\hbar\omega_{n0}=E_n-E_0$, $\tilde{H}|n\rangle= E_n |n\rangle$.
Further, $\epsilon_0$ is the contribution of all
the other high frequency modes, $V$ is the total volume.
The expression for the magnetic response function, $\mu_{ik}(\omega)$, is obtained by replacing
$\tilde{{\bf P}}$ with $g\mu_B{\bf S}$ (we neglect the effect of $\tilde{{\bf L}}$ relative
to the spin contribution and the difference between $\tilde{{\bf S}}$ and ${\bf S}$).

In absence of spin-orbit coupling, the states $|n\rangle$ have well defined total spin $S$ 
and $z$-projection $S_z$ (they are eigenstates of the isotropic Heisenberg Hamiltonian). The operator 
$\tilde{{\bf P}}$ preserves these quantum numbers. In contrast,  
$S_x$ and $S_y$ connect states with different total spin. 
Therefore, the excited states that contribute to $\epsilon_{ik}(\omega)$ and $\mu_{ik}(\omega)$ 
are in general different and their resonances are
different too. 
However, in the presence of spin-orbit coupling some resonances may be common for both as we will see below. 
It is important to note that the matrix elements of $\tilde{{\bf P}}$ are of order $8eat^3/U^3$, i.e., 
about the same order of magnitude as the matrix elements of $g\mu_B{\bf S}$ for $J\sim 100$ K. 
Here $a$ is a characteristic interatomic distance. 
Hence, the response of a Mott insulator to an ac electric field may be similar 
in magnitude to the response to an ac magnetic field. 
We note that if  $|0\rangle$ is an eigenstate of $\tilde{L}_z$ 
with nonzero eigenvalue (orbital currents), 
the matrix elements $\langle 0|\tilde{P}_x|0 \rangle $ and $\langle 0|\tilde{P}_y|0 \rangle $ 
are simultaneously nonzero leading to circular dichroism or rotation of the electric field polarization.
This rotation is 
almost the same as Faraday rotation induced by spins on the ac magnetic field polarization. 
Hence, detecting of orbital currents (``scalar spin chirality'') is possible 
by measuring the rotation of the electric field polarization. \cite{Varma}
To see that $\langle 0|\tilde{P}_i|n\rangle\neq 0$ for $i=x,y$ if there is a net orbital current in the state $|0\rangle$, we 
note that for electrons moving on a ring (or loop in general) the orbital moment $\tilde{L}_z$ is the conjugate 
variable of the angle $\varphi$ and  $\tilde{P}_x\propto \cos\varphi$, while $\tilde{P}_y\propto\sin\varphi$. 

The single equilateral triangle of $S=1/2$ spins with antiferromagnetic 
Heisenberg and spin-orbit interactions provides the
simplest realization of our results. There are many compounds, known as trinuclear spin complexes, 
that contain such isolated triangles. Triangular clusters exist in   
magnetic molecules like ``V15'' K$_6$[V$^{{\rm IV}}_{15}$As$_6$O$_{42}$(H$_2$O)] $\cdot$ 8H$_2$O \cite{Tsuker} 
or form well-ordered solids \cite{Broholm}. In the absence of Dzyaloshinsky-Moriya (DM) 
coupling, the ground state is a quartet with total spin $S=1/2$. The higher energy  
$S= 3/2$ quartet is separated from the ground state quartet by the gap $3J/2$.
The four lowest degenerate states, $\left\vert \chi,\sigma \right\rangle $, 
are labeled by ``spin chirality'' $\chi=\pm 1$ 
(which plays the role of the $z$--projection of a pseudospin variable ${\bf T}$) 
and spin projection of $S_{z}$ ($\sigma=\uparrow$ or $\downarrow$). The full space of the quartet can  
be presented as a direct product of spin and pseudospin subspaces. It turns out that $P_x\propto 3T_x=2{\bf S}_2\cdot{\bf S}_3-{\bf S}_1
\cdot({\bf S}_2+{\bf S}_3)$, $P_y\propto \sqrt{3}T_y= {\bf S}_1({\bf S}_2-{\bf S}_3)$ and 
and $L_z\propto \sqrt{3}T_z/2=\chi_{12,3}$, where the Pauli matrices ${\bf T}$ operate in the pseudospin subspace 
and obey SU(2) commutation relations. This is a consequence of the fact that current and polarization
are associated with conjugate variables (angular momentum and phase).
Thus matrix elements of both $P_x$ and $P_y$ are nonzero between states with opposite orbital momenta and spin chiralities.
Note that the eigenstates of $\tilde{L}_z$ break time reversal symmetry (see  Fig.\ref{triangle}a),
while eigenstates of $\tilde{P}_x\pm i\tilde{P}_y$ break the spatial $C_3$ symmetry (see Fig.\ref{triangle}b). 

\begin{figure}[!htb]
\vspace*{-0.3cm}
\hspace*{-0.4cm}
\includegraphics[angle=90,width=0.6\textwidth]{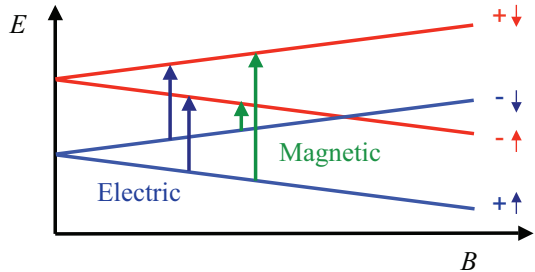}
\vspace{-4.8cm}
\caption{Energy levels of spins on triangles with the account of exchange, Dzyaloshinsky-Moriya and 
Zeeman interactions as a function of applied dc magnetic field. Blue (green) arrows show transitions 
induced by ac electric (magnetic, EPR) field.}
\label{esrlines}
\end{figure}

In some real systems like V15 (see Ref.~\onlinecite{Tsuker}, the lattice symmetry allows for 
a nonzero DM coupling
$H_{\mathrm{DM}}=\sum_{ij} {\bf D}_{ij}\cdot \lbrack \mathbf{S}_{i}\times 
\mathbf{S}_{j}]$. The terms that mix the $S=1/2$ and $S=3/2$ states 
(with in-plane components of the vector ${\bf D}_{ij}$) are
relatively small. On the other hand, the ${D}_{z}$ term 
plays the role of a spin-orbit coupling between the spin and the orbital 
moment $\tilde{L}_z$ and splits the ground state quartet into two doublets 
$\{| + ,\uparrow \rangle$, $| - ,\downarrow \rangle \}$
and $\{| + ,\downarrow \rangle$, $| - ,\uparrow \rangle\}$,
separated by an energy $\Delta=D_z$. 
Consequently, the system exhibits the following properties: 
I. In absence of a static magnetic field, the electron paramagnetic resonance (EPR) 
spectrum exhibits the same resonance frequency, $\omega_0=\Delta/\hbar$,  
as the dipole-allowed microwave absorption with similar intensities. 
In a static magnetic field, $B$, the absorption frequency due 
to the ac electric field remains the same while the EPR frequency splits linearly in $B$, see Fig.\ref{esrlines}. 
II. Slightly below $\omega_0$, the contribution of this resonance to both 
$\epsilon_{ii}(\omega)$ and $\mu_{ii}(\omega)$ for $i=x,y$ is negative and 
both will be negative if dissipation is weak (negative refraction index) \cite{Veselago}. 
III. Off-diagonal elements $\epsilon_{xy}(\omega)$ and $\epsilon_{yx}(\omega)$ are nonzero 
at low temperatures in the presence of the magnetic field which splits lowest doublet resulting in a
strong rotation of the electric field polarization at frequencies of order $\omega_0$. 
IV. The electric field causes transitions between states 
with the same total spin. Therefore, $\epsilon_{ik}(\omega)$ changes with the ground 
state magnetization until the contribution of magnetic states 
vanishes when all spins become aligned in the field direction.
Hence, measurements of the dielectric function, including the static case, 
provide information about the structure of magnetic spectrum.

While the states with nonzero orbital currents or polarization are degenerate in a single $C_3$-invariant triangle, 
this may not be the general case of infinitely large systems containing triangles in their
structures. Different symmetries can be broken in these systems.
If the resulting spin ordered state 
is such that $\langle \chi_{ij,k}\rangle \neq 0$, 
Eq.~(\ref{eq:current}) implies that the spin ordering is accompanied by an ordering of orbital moments.

In conclusion, magnetic states of Mott insulators show electrical properties which 
distinguish them from standard band insulators.
These states can exhibit electric dipole moments, 
which gives a purely electronic mechanism of multiferroic behavior. 
Spin states also contribute to the low-frequency optical 
properties, such as absorption by magnetic excitations well below the gap for single--electron excitations 
(Hubbard gap). An even more striking property of Mott insulators is the presence of orbital electronic currents 
and the corresponding orbital moments which
can be detected by measuring the resulting rotation of the electric field polarization or by nuclear magnetic resonance.

The authors thank A. Saxena for useful discussion. 
LANL is supported by US DOE under Contract No. W-7405-ENG-36. The work of D.Kh. 
was supported by the DFG via SFB 608 and the European project COMEPHS.


\begin{thebibliography}{99}
\bibitem{chirality}Actually $\chi_{12,3}$ characterizes the magnitude of an 
orbital current as shown in the paper and this object is achiral 
according to the strict mathematical definition of chirality.
\end{thebibliography}
\end{document}